\documentclass[ChapterTOCs,saunders1]{saunders} 
\usepackage{fixltx2e,fix-cm}
\usepackage{amssymb}
\usepackage{amsmath}
\usepackage{graphicx}
\usepackage{subfigure}
\usepackage{makeidx}
\usepackage{multicol}
\usepackage{mathptmx}
\usepackage{color}

\begin{document}

\frontmatter

\title{} 
\author{}

\tableofcontents
\listoffigures
\listoftables

\mainmatter

\chapterauthor{Michael Felderer}{University of Innsbruck, Austria}
\chapterauthor{Jeffrey C. Carver}{University of Alabama, USA}

\chapter{Guidelines for Systematic Mapping Studies in Security Engineering}

\section{Introduction} \label{sec:introduction} 
Maturation of a field requires that researchers be able to analyze and synthesize research to draw deeper, more meaningful conclusions.
As a research area matures there is often a sharp increase in the number of research reports and results made available.
With this increase, it is important to perform secondary studies that summarize results and provide an overview of the area. 
Methodologies exist for various type of secondary studies (i.e. systematic literature reviews and systematic mapping studies), which have been extensively used in evidence-based medicine and and software engineering. 
Secondary studies are less common in security engineering.
However, a general trend toward more empirical studies in software security and evidence-based software security engineering has lead to an increased focus on systematic research methods. 

Systematic mapping is a methodology that is frequently used in medical research and recently also in software engineering~\cite{petersen2015guidelines-mapping-studies-update,kitchenham2016evidencebasedse}, but largely neglected in (software) security engineering. 
\emph{Security engineering} focuses on security aspects in the software development lifecycle~\cite{felderer2014evolution}. 
Security engineering aims at protecting information and systems from unauthorized access, use, disclosure, disruption, modification, perusal, inspection, recording, or destruction.
It has a main objective of guaranteeing confidentiality, integrity, and availability of information and systems. 

A \emph{systematic mapping study} (SMS) provides a `map' of a research area by classifying papers and results based on relevant categories and counting the frequency of work in each of those categories.
Such a distribution provides an overview of a field to help research identified topics that are well-studied and topics that are in need of additional study.
Unlike Systematic Literature Reviews (SLRs), which seek to answer a specific research question based on all available evidence, SMSs are important because they (1) provide a basis for future research~\cite{kitchenham2011using} and (2) educate members of the community~\cite{kitchenham2010educational}.

Therefore, SMSs should be an important tool in a field like security engineering where the number of studies and the maturity of empirical methods used is increasing, but not yet mature enough for SLRs.
It is important for researchers to provide a map of key topics like security requirements engineering or security testing.
We identified only ten mapping studies published in the field of security engineering
\cite{arshad2011security,daSilva2013systematic,felderer2015regtest,carver2016establishing,felderer2016mbst,nunes2016proposal,sauerwein2016systematic,souag2016reusable}. 
We also note that there is no methodological support available that is specific to security engineering. 

The number of mapping studies is continuously increasing and there is a great interest in the methodology~\cite{petersen2015guidelines-mapping-studies-update}. 
To increase the confidence and reliability of mapping studies, there is a need for methodological support, especially domain-specific support. 
This need is illustrated in a recent analysis that showed large differences in terms of the included papers and the classification of results between two mapping studies on the same topic~\cite{wohlin2013reliability}.

This chapter provides methodological support for SMSs in security engineering based on examples from published SMSs in the field of security engineering and on our own experience with mapping studies.
Similar to software engineering also security engineering bridges research and practice and as the available SMSs in security engineering are similar to SMSs in software engineering, the same basic process for conducting mapping studies can be followed, with some tailoring for security engineering.
The SMS process consists of the following phases: 
(1) study planning, 
(2) searching for studies, 
(3) study selection, 
(4) assessment of study quality, 
(5) extraction of data,
(6) classification of data,
(7) analysis of data, and
(8) reporting of results~\cite{kitchenham2016evidencebasedse}.
The goal of this chapter is to build on this foundation with guidance that is specific to security engineering SMSs and to increase the awareness within the security engineering community of the need for additional SMSs on important topics.

This chapter is structured as follows. 
Section~\ref{sec:mapping-studies-software-eng} provides background on systematic mapping studies in software engineering.
Section~\ref{sec:mapping-studies-sec-eng} overviews the published security engineering SMSs. 
Section~\ref{sec:guidelines} presents guidelines for SMSs in security engineering aligned with the phases for conducting systematic mapping. 
Finally, Section~\ref{sec:summary} summarizes the chapter.

\section{Background on Systematic Mapping Studies in Software Engineering} \label{sec:mapping-studies-software-eng}

The goal of SMSs (which are also called scoping studies) is to survey the available knowledge about a topic~\cite{kitchenham2016evidencebasedse} to identify clusters of studies suitable for more detailed investigation and to identify gaps where more research is needed. 
While SMSs have found wide acceptance in software engineering, they appear to be less widely used in other disciplines~\cite{kitchenham2016evidencebasedse}. 
There use in software engineering is the result of its general immaturity as an `empirical' discipline.
That is, software engineering researchers do not yet use a consistent approach for reporting studies, nor have they agreed on a consistent set of terms to describe key concepts in the field.
SMSs are particularly applicable in this situation~\cite{kitchenham2007guidelines}.
We claim that security engineering is at a similar level of maturity as software engineering in regards to the reporting of empirical studies.
Therefore, SMSs should be of similar importance. 
But, unlike software engineering, there are no specific guidelines for performing SMSs in security engineering and the overall awareness of the importance of mapping studies within the community is lower. 
This chapter fills this gap and provides specific guidelines for performing SMSs in security engineering.
As a background, this section summarizes the relevant methodological support for SMSs in software engineering. 
Section~\ref{sec:se_process} describes the process for performing SMSs.
Section~\ref{sec:se_guidelines} discusses additional guidelines for SMSs in software engineering.

\subsection{Process for Systematic Mapping Studies in Software Engineering} \label{sec:se_process}

\begin{figure}[htb]
\centerline{\includegraphics[height=.5\textheight]{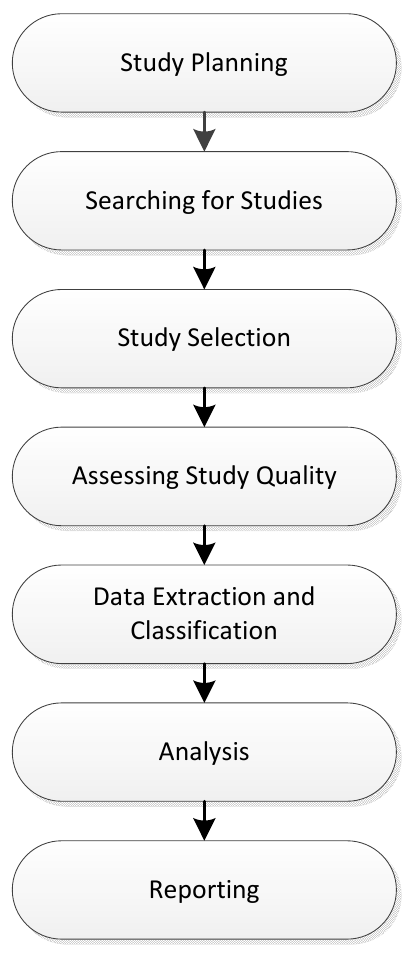}}
\caption{Process Steps for Systematic Mapping Studies.}
\label{fig:process}
\end{figure}

The SMS approach in software engineering is based on a process presented in a recent comprehensive book on evidence-based software engineering~\cite{kitchenham2016evidencebasedse}, which takes experiences from previous more generic~\cite{kitchenham2007guidelines} and more specific processes~\cite{petersen2008sysmappingse} into account. 
Figure~\ref{fig:process} shows the steps of the SMS process (i.e. study planning, searching for studies, study selection, assessing study quality, data extraction, data classification, analysis, and reporting).
We align the presentation of guidelines for SMSs in security engineering presented in Section~\ref{sec:guidelines} with these steps. 
Even though we discus the steps sequentially, in practice the steps are performed iteratively or even in parallel.
In addition to the SMS process described here, interested readers may want to review the SMS process described by Peterson et al.~\cite{petersen2008sysmappingse}, which is similar to the one described here.
The remainder of this section describes the SMS process steps in more detail.

\subsubsection{Study Planning} \label{sec:guidelines-planning}

As undertaking a mapping study is a time-consuming activity, planning is a key factor in achieving a successful outcome. 
The focus in this phase is on the development and validation of a review protocol. 
Prior to planning a full study, the researchers should first ensure that such an SMS is needed and feasible. 
SMSs require a justification and have to be managed as review projects. 
According to Kitchenham et al.~\cite{kitchenham2016evidencebasedse}, study planning includes (1) establishing the need for a review, (2) managing the review protocol, (3) specifying the research questions, (4) developing the review protocol, and (5) validating the review protocol. 

Specification of the research questions is a critical part of the planning process.
It is important for the researchers to fully explain the factors that motivate the research questions. 
For SMSs, the research questions should help the researchers~\cite{kitchenham2016evidencebasedse}:
\begin{itemize}
  \item classify the literature on a particular software engineering topic in ways that are interesting and useful to the community; and
  \item identify clusters of related research as motivation for further analysis and identify gaps in the literature which are in need of additional studies.
\end{itemize}

The main components of a review protocol aligned with the process steps are background, research questions, search strategy, study selection, assessing the quality of the primary studies, data extraction, analysis, limitations, reporting, and review management. 
Once developed, the protocol authors should provide protocol evaluators with a checklist or a set of questions addressing each of the elements of a protocol.
This validation process helps ensure the process is correct prior to embarking on the review.

\subsubsection{Searching for Studies} \label{sec:guidelines-searching}

There are a number of approaches researchers can use to identify relevant primary studies for inclusion in the SMS. 
These methods include:
\begin{itemize}
	\item \emph{Automated Searching} -- use resources like digital libraries and indexing systems;	
	\item \emph{Manual Searching} -- focusing on selected journals and conference proceedings;
	\item \emph{Backwards Snowballing} -- examining papers that are cited by other papers included in the review; and
	\item \emph{Forward Snowballing} -- examining papers that cite papers included in the review.
\end{itemize}
While, automate searching is probably the most common approach, in practice, researchers often combine these methods to develop a search strategy that achieves a good level of coverage for the topic of interest~\cite{kitchenham2016evidencebasedse}. 

There are a number of databases researchers use when performing SMSs.
First, there are publisher-specific databases including: \textit{IEEEXplore}, the \textit{ACM Digital Library}, \emph{Science Direct} (Elsevier), \emph{Wiley}, and \emph{SpringerLink}. 
Second, there are general indexing servcies, such as \textit{Scopus} and \emph{Web of Science} that index papers published by multiple publishers.
Finally, \textit{Google Scholar} can be useful, although there is less validation of paper prior to being indexed by Google Scholar.

The goal of the search process is to achieve an acceptable level of \emph{completeness} within the reviewer's constraints of time and human resources~\cite{kitchenham2016evidencebasedse}. 
Therefore developing a search strategy is an iterative process, involving refinement based on some determination of the level of completeness achieved. 
An essential basis for the subjective assessment of completeness is comparison of the search results against a \emph{known set} of papers.
This set contains papers for which the researcher has already determined their relevance to the current SMS.
Researchers can create the known set through an informal automated search, based upon their own personal knowledge, from a previous literature review, or through the construction of a \emph{quasi-gold standard} by performing a manual search across a limited set of topic-specific journals and conference proceedings over a restricted time period.
If the set of papers returned by the search string does not include papers in the \emph{known set}, then the researcher must modify the search string.

\subsubsection{Study Selection} \label{sec:guidelines-selection}

Study selection is a multi-stage process that can overlap to some extent with the searching process. 
The study selection process includes: determining the selection criteria, performing the selection process, and determining the and relationship between papers and studies. 

The \emph{selection criteria} are generally expressed as inclusion and exclusion criteria.
These criteria help the researcher make a more objective determination about which paper should be included in the review.
While the specific critera may vary depending upon the research questions, they often include items like: ensuring the paper is English, ensuring the paper is peer-reviewed, or ensuring the paper is within a specified time frame.
 
The \emph{selection process} usually consists of a number of stages. 
First, after researchers identify a set of candidate papers, they examine the titles and abstracts to eliminate any papers that are clearly irrelevant.
If there is a question about the relevance of a paper, then it remains for the next phase.
Next, researchers have to examine the papers in more detail to finally determine their relevance.
When study selection is performed by a team of reviewers, researchers can validate the results by performing the process independently and comparing the results.

Finally, determining the relationship between research papers and the studies that they report is important. 
Researchers performing SMSs are usually looking for empirical studies that provide some sort of evidence about a topic of interest. 
Papers can report more than one study, and studies can be reported in more than one paper.
Therefore, it is important for the researchers to eliminate duplicate reports and identify a set of unique studies from which data can be extracted.

\subsubsection{Assessing Study Quality} \label{sec:guidelines-assessment}

As the focus of an SMS is usually on classifying information or knowledge about a topic, quality assessment is of typically of less importance than for other quantitative or qualitative reviews. 
Researchers have to define criteria (for instance, quality checklists) and procedures for performing quality assessment (i.e. scoring studies, validating the scores as well as using quality assessment results).

\subsubsection{Data Extraction and Classification} \label{sec-guidelines-extraction}

The objective of this stage of the process is to analyze the reports of the primary studies to extract the data needed to address the research questions. 
Researchers must define, and justify, the strategy for data extraction and the data extraction form (which helps ensure researchers extract the same information about each study). 
With the resulting \emph{classification scheme}, the data extraction and aggregation may be performed iteratively, and it may be revised as more knowledge about the topic is gained through extraction and aggregation. 
Furthermore, the data extraction process should be validated.
Validation can be performed by having two independent researchers extract data from the same papers, checking the results, and meeting to reconcile any discrepancies.

\subsubsection{Analysis} \label{sec:guidelines-analysis}

The analysis of data gathered for an SMS is relatively simple compared to the analysis performed in a systematic literature review. 
In an SMS the data extracted from each primary study in a mapping study tends to be less detailed than the data extracted for a systematic literature review, in which a qualitative or quantitative analysis is common. 
The type of analyses typically performed in an SMS include: analysis of publication details, classification of results, and more complex automated content analysis. 

An \emph{analysis of publication details} can answer many research questions including: the distribution of authors, affiliations, publication dates, publication type, and publication venue.
A \emph{classification analysis} often produces more interesting results.
For example, this analysis may identify the existing techniques used in a particular area or the types of empirical methods used in studies.
Finally, \emph{automated content analysis} uses test mining and associated visualization methods to analyze mapping study data.

\subsubsection{Reporting} \label{sec:guidelines-reporting}

The first step in the reporting phase is \emph{planning} the report.
This step includes specifying the audience and determining what sort of document would best suit their needs. 
Ideally, report planning should be initiated during the preparation of a review protocol. 
The second step is \emph{writing} the actual report. 
Finally, the last step is \emph{validating} the reports through internal and/or external reviews to assess its quality.

\subsection{Overview of Guidelines for Systematic Mapping Studies in Software Engineering} \label{sec:se_guidelines}

Due to the importance of mapping studies in software engineering, there are several sets of guidelines published by different authors.
We can group the guidelines used for mapping studies in software engineering into four sets~\cite{petersen2015guidelines-mapping-studies-update}, as shown in Table~\ref{tab:guidelines-se}.

The first family is the \emph{Kitchenham family}.
This family consists of the generic guidelines on systematic reviews by Kitchenham and Charters~\cite{kitchenham2007guidelines}, which are a refinement of the original guidelines from Kitchenham~\cite{kitchenham2004procedures}.
It also includes a checklist for conducting mapping studies in software engineering (Template for a Mapping Study Protocol from Durham University)~\cite{durhamtemplate}.
These guidelines were refined in a later work by Kitchenham and Brereton~\cite{kitchenham2013systematic} and in a book by Kitchenham et al.~\cite{kitchenham2016evidencebasedse}.

The second family is the \emph{Petersen family}.
Petersen et al. performed a systematic map of 52 mapping studies in software engineering.
Based on this study, they produced evidence-based guidelines for systematic mapping studies in software engineering~\cite{petersen2015guidelines-mapping-studies-update}, which update previous guidelines~\cite{petersen2008sysmappingse} by the same first author.

The Kitchenham and Petersen families of guidelines are most commonly used~\cite{petersen2015guidelines-mapping-studies-update}.
The other two families, i.e. \emph{Bilolchini} and \emph{Social Science}, are less frequently used.

\begin{table}[hbt!]
\tabletitle{Guidelines used for Systematic Mapping Studies in Software Engineering}%
\label{tab:guidelines-se}
\begin{tabular}{p{2.5cm}p{4cm}p{4cm}}
\tch{Guideline Family} &\tch{References} \\
Kitchenham & \cite{kitchenham2016evidencebasedse}, \cite{kitchenham2013systematic}, \cite{kitchenham2007guidelines}, \cite{kitchenham2004procedures}, \cite{durhamtemplate} \\
Petersen & \cite{petersen2015guidelines-mapping-studies-update}, \cite{petersen2008sysmappingse} \\
Biolchini & \cite{biolchini2005systematic} \\
Social Science & \cite{petticrew2008systematic}, \cite{arksey2005scoping} \\
\end{tabular}
\end{table}

Each family of guidelines follows a process similar to the one presented in Section~\ref{sec:se_process}. The Kitchenham familiy provides the most comprehensive familiy of guidelines for systematic reviews in software engineering covering quantitative and qualitative reviews as well as systematic mapping studies. The Petersen familiy provides guidelines specific to systematic mapping studies, which are even based on an assessment of the current practice of conducting SMS by analyzing 52 SMS in software engineering. The focus of Biolchini is on providing a template for systematic literature reviews in software engineering, which can also be instantiated for SMSs. Finally, the systematic review guidelines for social sciences are more generic and neither cover specific issues of software engineering nor of mapping studies.  

Note that in software engineering, researchers often combine guidelines when performing SMSs because the individual guidelines are not sufficient~\cite{petersen2015guidelines-mapping-studies-update}. 
The same situation holds for security engineering, for which we provide a collection of guidelines in Section~\ref{sec:guidelines} of this chapter.

\section{Overview of Available Mapping Studies in Security Engineering} \label{sec:mapping-studies-sec-eng}

This section provides an overview of available mapping studies in security engineering. Mapping studies in security engineering provide a `map' of a security engineering research area by classifying papers and results based on relevant categories and counting the frequency of work in each of those categories. We do not consider systematic literature reviews in security enginering (which seek to answer a specific security engineering research question based on all available evidence), like the study by Oueslati et al. on the challenges of developing secure software using using the agile approach~\cite{oueslati2015literature}. 
Furthermore, we exclude those SMSs in software engineering which consider non-functional aspects including security as one classification item, for instance, the study by Afszal et al.\cite{afzal2008systematic,haser2014software}. 


\begin{table}[hbt!]
	\tabletitle{Security Engineering Mapping Studies}%
	\label{tab:mapping-studies-sec-eng}
	\begin{tabular}{lccc}
		\tch{Reference}    &\tch{Topic} &\tch{Year} &\tch{\# Venue\footnote{Abbreviations of venue names: CSI - Computer Standards \& Security, EASE - International Conference on Evaluation and Assessment in Software Engineering, IJCSIS - International Journal of Computer Science and Information Security, STTT - International Journal on Software Tools for Technology Transfer, HotSoS - Symposium and Bootcamp on the Science of Security, STVR - International Journal on Software Testing, Verification, and Reliability, IJCEACIE - International Journal of Computer, Electrical, Automation, Control and Information Engineering, SOSE - Symposium on Service-Oriented System Engineering, REJ - Requirements Engineering Journal, ASSD - Workshop on Agile Secure Software Development}} \\
		\cite{mellado2010systematic} & security requirements engineering & 2010 & CSI \\
		\cite{arshad2011security} & software architecture security & 2011   & EASE\\
		\cite{daSilva2013systematic} & security threats in cloud computing & 2013 & IJCSIS \\
		\cite{felderer2015regtest}   & security regression testing & 2015 & STTT \\
		\cite{carver2016establishing} & science of security & 2016 & HotSoS \\
		\cite{felderer2016mbst} & model-based security testing & 2016 & STVR \\
		\cite{nunes2016proposal} & security verification and validation & 2016 & IJCEACIE \\
		\cite{sauerwein2016systematic} & security crowdsourcing & 2016 & SOSE \\
		\cite{souag2016reusable} & security requirements engineering & 2016 & REJ \\
		\cite{Mohan2016SecDevOps} & security development and operations & 2016 & ASSD \\
	\end{tabular}
\end{table}

\begin{table}[hbt!]
\tabletitle{Number of Retrieved and Included Papers in Available Mapping Studies in Security Engineering.}%
\label{tab:mapping-studies-sec-eng-papers}
\begin{tabular}{lcccc}
\tch{Reference} &\tch{Year} &\tch{\# Retrieved Papers} &\tch{\# Included Papers} \\
\cite{mellado2010systematic} & 2010 & n/a & 21 \\
\cite{arshad2011security} & 2011 & 751 & 40 \\
\cite{daSilva2013systematic} & 2013 & 1011 & 661 \\
\cite{felderer2015regtest} & 2015 & 1095 & 18 \\
\cite{carver2016establishing} & 2016 & 55 & 55\footnote{This study provided a mapping of all papers published in the IEEE Symposium on Security and Privacy 2015 therefore the number of retrieved and included papers are equal.} \\
\cite{felderer2016mbst} & 2016 & 5928 & 119 \\
\cite{nunes2016proposal} & 2016 & 1216 & 55 \\
\cite{sauerwein2016systematic} & 2016 & 1535 & 23 \\
\cite{souag2016reusable} & 2016 & 158 & 95 \\
\cite{Mohan2016SecDevOps} & 2016 & 71 & 8 \\
\end{tabular}
\end{table}

Tables~\ref{tab:mapping-studies-sec-eng} and~\ref{tab:mapping-studies-sec-eng-papers} provide an overview of the published security engineering SMSs.
These SMSs cover different phases of the security development lifecycle~\cite{felderer2015process}: one SMS addresses security requirements engineering~\cite{souag2016reusable}, one security architectures~\cite{arshad2011security}, three security testing~\cite{felderer2015regtest,felderer2016mbst,nunes2016proposal}, one security deployment~\cite{daSilva2013systematic}, and one security development and operations~\cite{Mohan2016SecDevOps}. Finally, two SMS are more on the meta-level and cover the whole security development lifecyle~\cite{sauerwein2016systematic,carver2016establishing}.

In more detail, the following topics and classification criteria are applied in each of the available SMS in security engineering. 

Souag et al.~\cite{souag2016reusable} provide an SMS on reusable knowledge in security requirements engineering classifying papers according to the criteria knowledge reliance, knowledge representation forms, techniques for (re)using knowledge, tools for automation, as well as challenges regarding security knowledge (re)use in security requirements engineering. 

Arshad and Usman~\cite{arshad2011security} provide an SMS on security at software architecture level classifying papers according to bibliographic information.

Felderer and Fourneret~\cite{felderer2015regtest} provide an SMS on security regression testing classifying papers according to the abstraction level of testing, addressed security issues, the regression testing techniques, tool support, the evaluated system, the maturity of the evaluated system as well as evaluation measures.

Felderer et al.~\cite{felderer2016mbst} provide an SMS on model-based security testing classifying papers according to filter criteria (i.e. model of system security, model of the environment and explicit test selection criteria) as well as evidence criteria (i.e. maturity of evaluated system, evidence measures and evidence level).

Nunes and Abuquerque~\cite{nunes2016proposal} provide an SMS on software security verification, validation and testing (VVT) classifying papers according to covered VVT practices and their advantages and difficulties. 

da Silva et al.~\cite{daSilva2013systematic} provide an SMS on security threats in cloud computing classifying papers according to security threats to cloud computing models, security domains explored by the threats, types of proposed solutions, as well as involved compliances.

Mohan and ben Othmane~\cite{Mohan2016SecDevOps} provide an SMS on security aspects in development and operations (DevOps) classifying papers and presentations from selected security conferences according to the criteria definition, security best practices, compliance, process automation, tools for SecDevOps, software configuration, team collaboration, availability of activity data and information secrecy.

Sauerwein et al.~\cite{sauerwein2016systematic} provide an SMS on crowdsourcing in information security classifying papers according to bibliographic information, applied research methodology, addressed information security application context, applied crowdsourcing approach, and challenges for crowdsourcing-based research in information security.

Carver et al.~\cite{carver2016establishing} provide an SMS on science of security based on the 2015 IEEE Security \& Privacy Proceedings classifying papers according to evaluation subject type, newness, evaluation approach and completion rubrics (i.e. research objectives, subject/case selection, description of data collection procedures, description of data analysis procedures, as well as threats to validity).

In the next section, we substantiate the provided guidelines with examples from these systematic mapping studies on security engineering.

\section{Guidelines for Systematic Mapping Studies in Security Engineering} \label{sec:guidelines}

Section~\ref{sec:mapping-studies-software-eng} provided a general overview for conducting mapping studies.
This section focuses those guidelines into the software security engineering domain by providing specific examples from the series of mapping studies described in Section~\ref{sec:mapping-studies-sec-eng}.
Note that the papers did not report anything specific to security as it relates to \textit{Analysis} or \textit{Reporting}. 
Therefore, we do not include a discussion about those aspects of the process.

\subsection{Study Planning} \label{sec:guidelines-planning}
The most common recommendation from the security mapping studies that affects study planning was to keep the scope of the study as general as possible.
One way to keep the scope general is to ensure that the research question is general rather than being too specific~\cite{arshad2011security,nunes2016proposal}.
Some studies use the PICOC (P -- problem or population, I -- intervention, C -- comparison, control or comparator, O -- outcomes, C -- context) approach to define the research question~\cite{nunes2016proposal}.
In addition, the goal of a mapping study should be to provide an overview of a topic to 
(1) identify which sub-topics are ready for Systematic Literature Reviews and which need additional primary studies first, and
(2) provide a reference for PhD students and other researcher~\cite{souag2016reusable}.

Other considerations during the planning phase include:
(1) ensuring that the protocol is defined in a replicable manner~\cite{sauerwein2016systematic},
(2) choosing the tooling infrastructure, e.g. Mellando et al. recommend Endnote~\cite{mellado2010systematic}, and
(3) planning for potential bottlenecks and sources of bias, e.g. Mellando et al. found that searching for, selecting, and evaluating studies led to bottlenecks and potential sources of bias~\cite{mellado2010systematic}.

\subsection{Searching for Studies} \label{sec:guidelines-searching}
To provide some guidance on the most important venues in which security engineering papers are published, we examined the `Computer Security \& Cryptography' list from Google Scholar and the `Computer Security' list from Microsoft Academic Search to identify the main journals and conferences related to security engineering.
To keep the list relevant, we did not consider venues focused information security management, cryptology, trust, computer security or network security. 
Table~\ref{tab:journals-security} and Table~\ref{tab:conf-security} list the main journals and conferences, respectively.
The `Studies Included' column in each table indicates whether any of the mapping studies described in Section~\ref{sec:mapping-studies-sec-eng} included a primary study from that venue.

\begin{table}[htb!]
\tabletitle{Main Security Journals Covering Software Security Engineering Topics.}%
\label{tab:journals-security}
\begin{tabular}{lc}
\tch{Journal Name} & \tch{Studies Included} \\ 
Computers \& Security & no \\
Information Security Journal: A Global Perspective & yes  \\
International Journal of Information Security & yes \\
Journal of Information Security and Applications & no \\
Security \& Privacy & yes \\
Transactions on Dependable and Secure Computing & yes \\
Transactions on Information Forensics and Security & no \\
Transactions on Privacy and Security & yes \\ 
\end{tabular}
\end{table}

\begin{table}[htb!]
\tabletitle{Main Security-related Conferences Covering Software Security Engineering Topics.}%
\label{tab:conf-security}
\begin{tabular}{lc}
\tch{Conference Name} & \tch{Studies Included} \\
Annual Computer Security \\\hspace{.2cm} Applications Conference (ACSAC) & yes \\
European Symposium on \\\hspace{.2cm} Research in Computer Security (ESORICS) & yes \\
International Conference on \\\hspace{.2cm} Availability, Reliability and Security (ARES) & yes \\
International Conference on \\\hspace{.2cm} Software Quality, Reliability \& Security (QRS) & yes \\
International Symposium on \\\hspace{.2cm} Engineering Secure Software and Systems (ESSoS) & yes \\
Symposium on \\\hspace{.2cm} Access Control Models and Technologies (SACMAT) & yes \\
Symposium on \\\hspace{.2cm} Computer and Communications Security (CCS) & yes \\
Symposium on \\\hspace{.2cm} Security and Privacy & yes \\
USENIX Security Symposium & yes \\
\end{tabular}
\end{table}

We also note that security engineering topics appear in domain specific venues. 
Specifically, there are a number of software engineering journals and conferences that cover security engineering topics.
The security engineering mapping studies in Table~\ref{tab:mapping-studies-sec-eng} include primary studies from the main generic software engineering venues listed in Table~\ref{tab:venues-se}. 

\begin{table}[htb!]
\tabletitle{Main Software Engineering Venues Covered by Security Engineering Mapping Studies.}%
\label{tab:venues-se}
\begin{tabular}{lc}
\tch{Journal Name} & \tch{Studies Included} \\ 
Transaction on \\\hspace{.2cm} Software Engineering (TSE) & yes \\
Transactions on \\\hspace{.2cm} Software Engineering and Methodology (TOSEM) & yes  \\
Journal of \\\hspace{.2cm} Systems and Software (JSS) & yes \\
Information and \\\hspace{.2cm} Software Technology (IST) & yes \\
International Conference on \\\hspace{.2cm} Software Engineering (ICSE) & yes \\
International Symposium on \\\hspace{.2cm} the Foundations of Software Engineering (FSE) & yes \\
International Conference on \\\hspace{.2cm} Automated Software Engineering (ASE) & yes \\
\end{tabular}
\end{table}

Furthermore, security mapping studies that focus on specific sub-topics may also find primary studies in the leading venues from that sub-area. 
For instance, the mapping study of Mellado et al.~\cite{mellado2010systematic} on security requirements engineering covers publications from the \textit{Requirements Engineering Journal (REJ)}, the \textit{International Requirements Engineering Conference (RE)} and the \textit{International Working Conference on Requirements Engineering: Foundation for Software Quality (REFSQ)}.

In another example Mohan and Othmane~\cite{Mohan2016SecDevOps} also searched for presentations at the OWASP\footnote{http://www.owasp.org} and RSA\footnote{https://www.rsaconference.com/} conferences. 
These two conferences are industry events that contain grey literature like presentations or blog entries.
For some recent or industry-relevant topics, the inclusion of grey literature in mapping studies could be quite useful~\cite{garousi2016need}. For instance, for the recent and industry-relevant topic of security aspects in DevOps covered by~\cite{Mohan2016SecDevOps}, the authors found 5 presentations (besides 66 research publications), and finally selected 3 presentations and 5 research papers for inclusion in the map.  

When performing the search process, it is important to have the most appropriate search string.
Similar to the discussion about the research question, using a broad search string will help prevent premature elimination of potentially relevant studies~\cite{daSilva2013systematic}.
To ensure the best coverage of the domain, authors have to choose an appropriate set of synonyms for the search string.
Arshad and Usman identified the following synonyms for \textit{security}: `Secure', `Authorization', `Authentication', and `Access Control'~\cite{arshad2011security}.

As is common with most systematic literature studies, authors have to use different search strings depending upon the venue being searched~\cite{souag2016reusable}. To provide some more specific guidance, Table~\ref{tab:databases} provides an overview of the literature databases searched in the identified software security mapping studies. Each of the ten security mapping studies searched in IEEE Xplore and the ACM Digital Library, which are also the two most common digital libraries used in software engineering mapping studies. Furthermore, more than half of the studies searched in ScienceDirect (eight out of ten) and in SpringerLink (five out of ten). The general indexing services Scopus, Compendex and Google Scholar are used in more than one study. The remaining databases, i.e. Engineering Village, Wiley, Web of Science, Citeseer, Taylor \& Francis, Wiley, and DBLP are each searched in one study.

\begin{table}[htb!]
	\tabletitle{Common Databases Used for Software Security Engineering Mapping Studies}
	\label{tab:databases}
	\begin{tabular}{l l}
		\textbf{Database} & \textbf{Studies Used}\\
		IEEE Xplore & \cite{arshad2011security,daSilva2013systematic,felderer2016mbst,felderer2015regtest,mellado2010systematic,nunes2016proposal,sauerwein2016systematic,souag2016reusable,Mohan2016SecDevOps} \\
		ACM Digital Library & \cite{arshad2011security,daSilva2013systematic,felderer2016mbst,felderer2015regtest,mellado2010systematic,nunes2016proposal,sauerwein2016systematic,souag2016reusable}\\
		ScienceDirect & \cite{arshad2011security,daSilva2013systematic,felderer2016mbst,felderer2015regtest,mellado2010systematic,nunes2016proposal,sauerwein2016systematic,souag2016reusable} \\
		SpringerLink & \cite{daSilva2013systematic,felderer2016mbst,felderer2015regtest,sauerwein2016systematic,souag2016reusable} \\
		Scopus & \cite{daSilva2013systematic,nunes2016proposal,sauerwein2016systematic} \\
		Compendex  & \cite{arshad2011security,nunes2016proposal}\\
		Google Scholar & \cite{mellado2010systematic,sauerwein2016systematic,Mohan2016SecDevOps}\\
		Engineering Village & \cite{daSilva2013systematic} \\
		Wiley  & \cite{felderer2016mbst} \\
		Web of Science & \cite{nunes2016proposal} \\
		Citeseer & \cite{sauerwein2016systematic} \\
		Taylor \& Francis & \cite{sauerwein2016systematic} \\
		Wiley & \cite{sauerwein2016systematic} \\		
		DBLP & \cite{souag2016reusable} \\
	\end{tabular}
\end{table}

Finally, it is important to pilot test the search string to ensure that it is providing the proper set of papers. 
Felderer et al. used a reference set of papers (which were known to be relevant) to evaluate the completeness of the set of papers returned by the search string~\cite{felderer2016mbst,felderer2015regtest}.

Another approach that augments that traditional search is \textit{snowballing}.
In this approach, researchers start from a set of papers that are known to be relevant.
This set can either come from prior work in the area or can be the result of the searching phase of the systematic mapping study.
For each paper in this set, the researcher check the list of references to determine if there are any additional papers that are not yet included.
If the authors find another relevant paper, they add it to the set and continue the snowballing process by checking its reference list.
We found two security mapping studies that employed the snowballing approach~\cite{sauerwein2016systematic,nunes2016proposal}.

\subsection{Study Selection} \label{sec:guidelines-selection}
Once authors identify an initial set of papers based on the search process described in Section~\ref{sec:guidelines-searching}, they must then choose the most appropriate set of papers to finally include in the mapping study.
Due to the subjective nature of this step and the potential for errors, a number of studies recommend using at least two members of the author team to perform the selection process.
Those authors should be experts in the domain~\cite{felderer2015regtest} and ideally have their work validated by other authors from the author team~\cite{souag2016reusable}.

The process of choosing the most appropriate studies would be prohibitively time-consuming if the authors needed to read the full text of all candidate papers.
Experiences from the security mapping studies provide some approaches to speed up this process.
Authors should start by examining the abstract to determine relevance.
When the abstract does not provide enough information to make a decision, there are different approaches.
First, Arshad and Usman recommend that authors next study the introduction and conclusion, then the full paper~\cite{arshad2011security}.
Conversely, Nunes and Albuquerque indicate that the abstract, introduction, and conclusion do not always contain enough information to make a decision~\cite{nunes2016proposal}.
Second, Saurwein et al. suggest that if a paper seems relevant, then perform partial reading of the paper~\cite{sauerwein2016systematic}.
Third, da Silva et al. point out that a full reading of the paper will be required if either (1) the abstract is too short or (2) the proposed solution is not fully described in the abstract~\cite{daSilva2013systematic}.
Finally, in some cases the information necessary to make a selection decision or to gather evidence may be completely omitted from paper and only located in an external thesis/dissertation~\cite{nunes2016proposal}.

A key part of the selection process is defining appropriate inclusion and exclusion criteria.
Here we provide some of the common criteria used by the authors of the included papers.

Common \textit{inclusion criteria} include:
\begin{itemize}
	\item Papers directly related to topic of study, including security and the specific focus~\cite{arshad2011security,daSilva2013systematic,felderer2016mbst,felderer2015regtest,nunes2016proposal,sauerwein2016systematic,Mohan2016SecDevOps}
	\item Published in a peer reviewed conference/journal~\cite{arshad2011security,felderer2016mbst,felderer2015regtest,nunes2016proposal,sauerwein2016systematic,Mohan2016SecDevOps}
	\item Paper should have a proposed solution or detailed description of process~\cite{daSilva2013systematic,mellado2010systematic,Mohan2016SecDevOps}
	\item Studies must be recent or withing a specified timeframe~\cite{mellado2010systematic,sauerwein2016systematic}
\end{itemize}

In addition to the negation of the inclusion criteria above, the following \textit{exclusion criteria} include:
\begin{itemize}
	\item Duplicate papers or older versions of current work~\cite{daSilva2013systematic,nunes2016proposal,sauerwein2016systematic}
	\item Papers not accessible online~\cite{daSilva2013systematic}
\end{itemize}

\subsection{Assessing Study Quality}
Only one paper specifically discussed how to assess the quality of the included studies.
Souag et al. provided four quality criteria used in their review to help determine whether to include a study~\cite{souag2016reusable}.
These criteria were:
\begin{itemize}
	\item Ensuring completeness of publication, including whether claims were supported and whether external work was cited
	\item Ensuring that the topic of the paper is closely related to the specific focus of the review;
	\item Grouping of similar work from sets of authors;  and
	\item Exclusion of least relevant articles to keep paper size manageable.
\end{itemize}

Related to ensuring the completeness of publications, Carver et al. developed a set of rubrics to help reviewers evaluate whether authors documented all key details about a study~\cite{carver2016establishing}.
Those rubrics include specific guidance related to:

\begin{itemize}
	\item \emph{Research objectives} -  are important to understand the goals of a paper and position the results;
	\item \emph{Subject/case selection} - help to understand how to interpret the results if the authors have clearly and explicitly described the subjects of the evaluation (e.g. the system or people chosen to participate), why those subjects are appropriate and how they were recruited or developed;
	\item \emph{Description of data collection procedures} - help to clarify exactly what information was collected;
	\item \emph{Description of data analysis procedures} - are important to enable replication;
	\item \emph{Threats to validity} - help to understand the limitations of the results and whether or not those results are applicable in a particular situation.
\end{itemize}

These rubrics are especially important to assess the quality of reporting of the provided evaluation and whether required data for classification of the provided evaluation can be extracted from a study. Furthermore, they help to determine whether other researchers are able to understand, replicate, and build-on research published in the respective study.

\subsection{Data Extraction and Classification} \label{sec-guidelines-extraction}
The identified mapping studies rarely discuss data extraction explicitly.
In fact, only half of the identified studies do not discuss data extraction at all \cite{arshad2011security,nunes2016proposal,daSilva2013systematic,felderer2015regtest,felderer2016mbst}. 
Three studies that do describe data extraction include nothing out of the ordinary~\cite{sauerwein2016systematic,mellado2010systematic,souag2016reusable}. 
Only two studies~\cite{souag2016reusable,Mohan2016SecDevOps} explicitly provide the information collected in the data extraction form.

Security is a diverse field in which researchers propose and evaluate a variety of artifacts.
Carver et al. identified six types of subjects that appear in the security engineering literature~\cite{carver2016establishing}, including:
\begin{itemize}
	\item \emph{Algorithm/theory} -  a proposal of a new algorithm/theory or an update to an existing algorithm/theory;
	\item \emph{Model} - a graphical or mathematical description of a system and/or its properties;
	\item \emph{Language} - a new programming language;
	\item \emph{Protocol} - a written procedural method that specifies the behavior for data exchange amongst multiple parties; 
	\item \emph{Process} - the computational steps required to transform one thing into something else;
	\item \emph{Tool} - an implementation of a process.
\end{itemize}

In terms of types of papers, Souag et al.~\cite{souag2016reusable} built a classification scheme for study types that is based on an established classification scheme used in requirements engineering~\cite{wieringa2006requirements}.
They classify papers as:
\begin{itemize} 
	\item \emph{solution proposals} -- propose a solution and argue for its relevance without a full-blown validation,
	\item \emph{philosophical papers} -- sketch a new way of looking at things, 
	\item \emph{evaluation research} -- investigate a problem in practice or an implementation of a technique in practice, 
	\item \emph{validation research} -- investigate the properties of a proposed solution that has not yet been implemented in practice,
	\item \emph{opinion papers} -- contain the author's subjective view on a topic, and 
	\item \emph{personal experience papers} -- emphasize the `what' rather than the `why'.
\end{itemize}

For the papers that fall into the \textit{evaluation research} or \textit{validation research} types, Carver et al. also provide a classification of the types of validation researchers may use~\cite{carver2016establishing}, including:
\begin{itemize}
	\item \emph{Experiment} -  an orderly process that seeks to test a hypothesis in a controlled setting to establish a causal relationship;
	\item \emph{Case study} - evaluation conducted in a more realistic setting that can be provided by an experiment;
	\item \emph{Survey} - comprises a number of qualitative research methods including surveys, interviews, focus groups, and opinion polls, which focus on gathering qualitative data (along with some quantitative data) directly from a set of subjects via a series of questions or guided discussion;
	\item \emph{Proof} - a formal approach to validate a characteristic or property of an evaluation subject; 
	\item \emph{Discussion/argument} - validation without empirical data, but through discussion or argument (this does not refer to papers that have a discussion of the results obtained by one of the other evaluation approaches).	
\end{itemize}

\begin{figure}[htb!]
\centerline{\includegraphics[width=\textwidth]{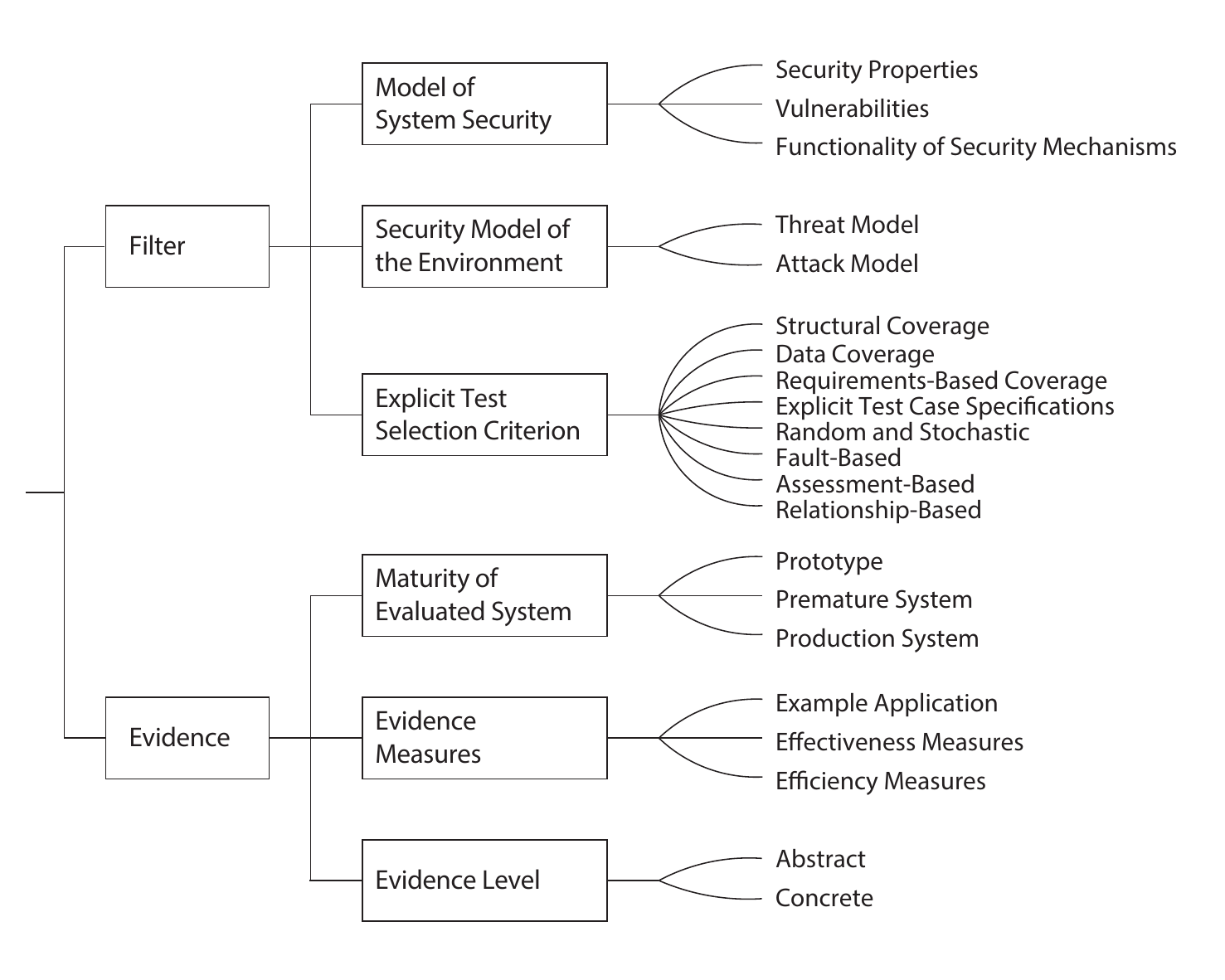}}
\caption{Classification Scheme for Model-Based Security Testing (from the Mapping Study Presented in~\cite{felderer2016mbst}).}
\label{fig:mbst-classification}
\end{figure}

A crucial artifact for data extraction and classification is the applied classification scheme, which may be refined during the process. Mapping studies in security engineering can \emph{complement classification schemes from available baseline studies in software engineering}. This provides the advantage for mapping studies in security engineering to base them on established classification schemes, to extend them with specific security aspects and to compare results to baseline results in software engineering. For instance, the classification scheme on model-based security testing from the mapping study in~\cite{felderer2016mbst}, which is shown in Figure~\ref{fig:mbst-classification}, complements a previously developed classification scheme for model-based testing~\cite{utting2012taxonomy}. 

This classification scheme for model-based testing comprises the top-level classes model specification, test generation and test execution. It is complemented in the mapping study on model-based security testing presented in~\cite{felderer2016mbst} with the class filter, which covers security aspects of model-based testing, i.e. model of system security, security model of the environment as well as explicit test selection criterion. Furthermore, the class evidence is added which considers the maturity of the evaluated system, evidence measures as well as the evidence level, which are are especially of importance for security testing approaches. However, the resulting classification scheme (with top-level classes model specification, test generation, test execution, filter, and evidence) extends the classification of model-based testing with classes for security testing and allows a refined classification of model-based security testing approaches as well as their comparison to model-based testing approaches.

\section{Summary} \label{sec:summary}

This chapter provides guidelines for SMSs in security engineering based upon specific examples drawn from published security engineering SMSs.
The goal of this chapter is to use these examples to provide guidelines for SMSs that are specific for security engineering and to increase the awareness in the security engineering community of the need of more SMSs. 
We present guidelines presented for the SMS phases study planning, searching for studies, study selection, assessing study quality, as well as data extraction and classification.
Although SMSs are important for the field of security engineering as a basis for future research and to stabilize terminology, they are still rare. The authors of this chapter hope that security engineering researchers find the provided guidelines useful and supportive to do more SMSs in the field.

\bibliographystyle{plain}
\bibliography{references}

\end{document}